\documentclass[sigconf, authorversion]{acmart}

\AtBeginDocument{%
  \providecommand\BibTeX{{%
    \normalfont B\kern-0.5em{\scshape i\kern-0.25em b}\kern-0.8em\TeX}}}

\usepackage{subcaption}
\usepackage{tabularx}
\usepackage{caption}
\usepackage{tabu}
\usepackage{booktabs}
\usepackage{float}
\usepackage{footnote}

\newcolumntype{P}[1]{>{\centering\arraybackslash}p{#1}}

\begin{document}

\title[Effects of Clutter on Egocentric Distance Perception in Virtual Reality]{Effects of Clutter on Egocentric Distance Perception in Virtual Reality}

\author{Sina Masnadi}
\email{sina@knights.ucf.edu}
\orcid{0000-0001-9874-8225}
\affiliation{%
  \institution{University of Central Florida}
  \streetaddress{4000 Central Florida Blvd}
  \city{Orlando}
  \state{FL}
  \country{USA}
  \postcode{32816}
}

\author{Yahya Hmaiti}
\email{yohanhmaiti@knights.ucf.edu}
\orcid{0000-0003-1052-1152}
\affiliation{%
  \institution{University of Central Florida}
  \streetaddress{4000 Central Florida Blvd}
  \city{Orlando}
  \state{FL}
  \country{USA}
  \postcode{32816}
}

\author{Eugene Taranta}
\email{etaranta@gmail.com}
\orcid{0000-0001-5615-2597}
\affiliation{%
  \institution{University of Central Florida}
  \streetaddress{4000 Central Florida Blvd}
  \city{Orlando}
  \state{FL}
  \country{USA}
  \postcode{32816}
}

\author{Joseph J. LaViola Jr.}
\email{jjl@cs.ucf.edu}
\orcid{0000-0003-1186-4130}
\affiliation{%
  \institution{University of Central Florida}
  \streetaddress{4000 Central Florida Blvd}
  \city{Orlando}
  \state{FL}
  \country{USA}
  \postcode{32816}
}

\renewcommand{\shortauthors}{Masnadi et al.}
\begin{abstract}
To assess the impact of clutter on egocentric distance perception, we performed a mixed-design study with 60 participants in four different virtual environments (VEs) with three levels of clutter. Additionally, we compared the indoor/outdoor VE characteristics and the HMD's FOV.
The participants wore a backpack computer and a wide FOV head-mounted display (HMD) as they blind-walked towards three distinct targets at distances of 3m, 4.5m, and 6m. The HMD's field of view (FOV) was programmatically limited to 165°$\times$110°, 110°$\times$110°, or 45°$\times$35°.
The results showed that increased clutter in the environment led to more precise distance judgment and less underestimation, independent of the FOV.
In comparison to outdoor VEs, indoor VEs showed more accurate distance judgment.
Additionally, participants made more accurate judgements while looking at the VEs through wider FOVs.

\end{abstract}
\maketitle


\section{Introduction}
Virtual Reality (VR) provides a unique experience for users where they are in a real place while interacting with a Virtual Environment and its components. Recently, VR technologies are becoming more user-friendly and widely available and immersive VEs using HMDs have shown increasing potential for productivity and entertainment. In VEs, the computer-generated 3D graphics envelops a user, and a tracking system is used to update the user’s viewpoint as they move.

The perception of distance in VEs is a long-studied topic in research related to VR interactions. 
Most VR applications and games try to make the user experience in their respective VEs resemble the real environment experience, which includes having smooth movements, natural behavior, interacting with the surrounding environment, etc. This highlights the importance of having an accurate representation of a VE. Due to the importance of precise representation of a VE in order to maintain and promote innate interactions, such as running, walking, and interacting with the surroundings, it is crucial to have an accurate egocentric distance perception. 

Numerous investigations have shed light on the fact that users often underestimate distances in VEs compared to the real-world~\cite{willemsen2009effects, geuss2010canIPass, mohler2006influence, pfeil2021distance, thompson2004does,renner2013review}. However, the factors that are causing this problem still need further investigation. In this paper, in order to better understand the conditions that might affect the perception of distance, we focused on the clutter in the environment as well as the FOV of the headset.

Our goal was to understand the potential influence of varying levels of clutter in an environment on egocentric distance judgment. The finding of such study a would benefit VE designers with the necessary information to make better decisions while designing VEs for distance-critical tasks. We hypothesized that increasing the clutter level in a given environment results in improved distance perception and less underestimation. 

Furthermore, recent work with the Pimax 5K HMD, the headset used in our study, provided good motivation for our comparison of clutter levels. In a comparison of a narrower FOV VR system versus modern wide-FOV systems, Masnadi et al. found that blind-walking performance, which consists of the user viewing a set target in the VE and then walking to it eyes closed, is more accurate in wide-FOV headsets compared to narrower FOVs~\cite{masnadi2022effects}. This motivated us to investigate if the FOV combined with clutter level has an effect on the perception of distance. In other words, we wanted to understand if FOV and clutter independently affect the perception of distance. Through our investigation, we found that the abundance and presence of clutter in the environment result in more accurate distance judgment and less underestimation.
Our work provides the following contributions to the literature on distance judgment and distance underestimation in VR:

\begin{itemize}
    \item A user study protocol that isolates clutter and FOV for direct evaluation in VEs.
    \item Evidence that clutter and FOV are important factors for reducing distance underestimation, and improving distance judgement in VEs.
    \item Evidence that the increase of clutter provides significant visual anchors that reduce distance underestimation.
    \item Evidence that FOV and clutter have independent effect on perception of distance.
\end{itemize}

In the rest of the paper, we first elaborate related work and compare our work with distance judgment and distance underestimation in VR literature. Afterwards, we report our user study in depth. Then, we proceed to report the results and after that we discuss the implications of our findings. At the end, we shed light on new considerations for VE design along with discussing the limitations and future work.

\section{Background and Related Work}

The distance underestimation phenomenon is due to different factors that have been investigated by prior research spanning the past two decades. These factors can be categorized into three main groups: virtual environment characteristics, apparatus characteristics, and evaluation techniques.

\subsection{Virtual Environment Characteristics}
Several research investigations on distance perception in VEs have discovered numerous factors that influence distance judgment in those environments. Kunz et al. showed that distance perception was influenced by the caliber and extent of the graphics quality related to the  VE~\cite{kunz2009revisiting}. Furthermore, the camera placement was shown by Leyrer et al. to contribute to distance underestimation, especially when the camera is positioned higher than the user's real eye-height~\cite{leyrer2011influence}. In our study, the camera was placed at the user's eye-level and the environments were designed to be detailed and realistic.

Aseeri et al. conducted experiments that showed the virtual human entourage (existence of life size human avatars in the scene) does not have a significant influence on the sense of scale in an indoor environment~\cite{aseeri2019investigating}. Self-avatars, when enabled in a  VE, contribute to better distance perception by the embodied users as shown by Mohler et al. ~\cite{mohler2010effectSelfAvatar}. In our study, we did not include a self-avatar to focus only on the effect of clutter on distance perception. 

The levels of detail in the environment and the visual cues available to the user can also affect distance and dimension judgment. 
Loyola et al. suggest that more available visual cues in a VE bring about a better accuracy in distance judgment and in dimension estimations, especially regarding egocentric dimensions~\cite{loyola2018influence}. Lessels et al. showed that having high-fidelity scenes in VEs is important and having a set simple control system leads to the preserving of spatial orientation, especially when fetching a target~\cite{lessels2005movement}. 

It has been shown by multiple studies that being in an indoor environment provides a better understanding of distances to the user when compared to an outdoor environment.
Creem-Regehr et al. performed a study which showed less underestimations in indoor VEs in contrast with outdoor VEs~\cite{creem2015egocentric}. In a related study, Houck et al. found that there is a robust influence of the width of indoor environment on egocentric distance perception which provides better understanding on what elements in an indoor environment is important~\cite{houck2022environment}. In addition, Andre and Rogers proved through a blind-walking evaluation that people underestimate distances greater in outdoor real-world environments compared to indoor environments~\cite{andre2006using}. Masnadi et al. also found that people tend to underestimate distances greater in outdoor VEs compared to indoor VEs~\cite{masnadi2022effects}. Our study is different from previous work since it evaluates multiple levels of clutter in various indoor and outdoor VEs. We focused on the effects of clutter as a the main factor in the misjudgment of egocentric distance in VR. With the abundance of visible objects, more visual cues are present, which might lead to the augmentation of perceivable objects by users. Thus, users have to look at more objects, ignore some or deviate their focus to others depending on their needs, which results in several effects on the visual perception of the user, considering that in the presence of visual clutter there is an increase in visual stimuli.



\subsection{Apparatus Characteristics}
It is also important to state the effects that the chosen apparatus might have on egocentric distance judgment. Part of a study lead by Combe et al. has shown that the weight of the HMD does not have a major effect over short distance perception~\cite{Combe9419335}. 
However, it was shown through several studies that the weight of an HMD and its inertia both influence the perception of distance in VR, with a tendency towards underestimation~\cite{buck2018comparison, jones2018degraded, willemsen2004effects}. 
Multiple investigations have shown that the judged distance accuracy was positively related to HMD FOV and its resolution, and negatively related to the HMD weight~\cite{kelly2022distance, masnadi2021field, masnadi2022effects}. 
Moreover, the parallax effect was shown not to impact the distance perception task of blind-walking~\cite{jones2008effects}. 
In our study, participants looked at the environment while standing in the same place, nevertheless, subtle moves could possibly happen before the start of the evaluation task. 

Pfeil et al. showed that distance underestimation remains a problem in video see-through HMDs and their results showed the influence of weight (negative correlation) and FOV of the HMD (positive correlation) on distance perception accuracy~\cite{pfeil2021distance}. Vaziri et al. reported similar results~\cite{vaziri2017impact, vaziri2021egocentric}.



\subsection{Evaluation Technique}
There are various evaluation techniques for analyzing the egocentric distance judgment in VEs. 

Some of the most popular techniques are blind-throwing, timed imagined walking, verbal estimation, and blind-walking~\cite{renner2013review,kitamoto4156284study}. In a \textit{blind throwing} task, users are blindfolded after they have seen the target in the environment and are ready to throw an object to the marked target~\cite{sahm2005throwing, ragan2012effects}. It has been shown in the case of \textit{verbal estimation}, the margin of error increases with the growth of the distance between the user and target~\cite{renner2013review, kunz2009revisiting, loomis2008measuring}. \textit{Timed imagined walking} consists of showing the target to the user, then when they are ready they are blindfolded and asked to imagine themselves walking to the target such that when they arrive there in their mind, they inform the investigator~\cite{grechkin2010howDoesPresentation, plumert2005distance}. 

In our study, we used the \textit{blind-walking} technique, which is the most popular method. It is similar to timed imagined walking, yet the difference is that the user physically walks towards the target while keeping their eyes closed or blind-folded, then they stop and let the researcher know they reached the target~\cite{willemsen2009effects, geuss2010canIPass, mohler2006influence, thompson2004does, masnadi2022effects}. 

\section{Methods}
We conducted a user study to evaluate distance judgment with different clutter levels in various VEs along with multiple HMD FOVs. The methods used for the study and their details are described in the sections that follow.

\subsection{Study Design}
The user study revolved around the blind-walking-based task. The task, which was set to estimate the user's perception of distance, consisted in asking the user to look at the VE through the HMD to determine the distance of a predefined target from themselves, then closing their eyes and walking to the target without opening their eyes during the walking process. 
We performed a $4 \times 3 \times 3 \times 3$ mixed-design study, such that the between-subject factors were environment-based characteristics - \textsc{two indoor} and \textsc{two outdoor} (4 levels) - along with the within-subject factors being the \textsc{clutter-level} (3 levels), \textsc{fov} (3 levels), and \textsc{target distance} (3 levels). In addition, the clutter levels, target distances, and FOV Dimensions used were consistent and the same across all the environments used in the study either indoors or outdoors.

\subsection{Study Variables}
\subsubsection{Between-Subjects Variables}
\begin{figure}
\centering
\subfloat[Indoor1 cluttered]{\includegraphics[width=\linewidth]{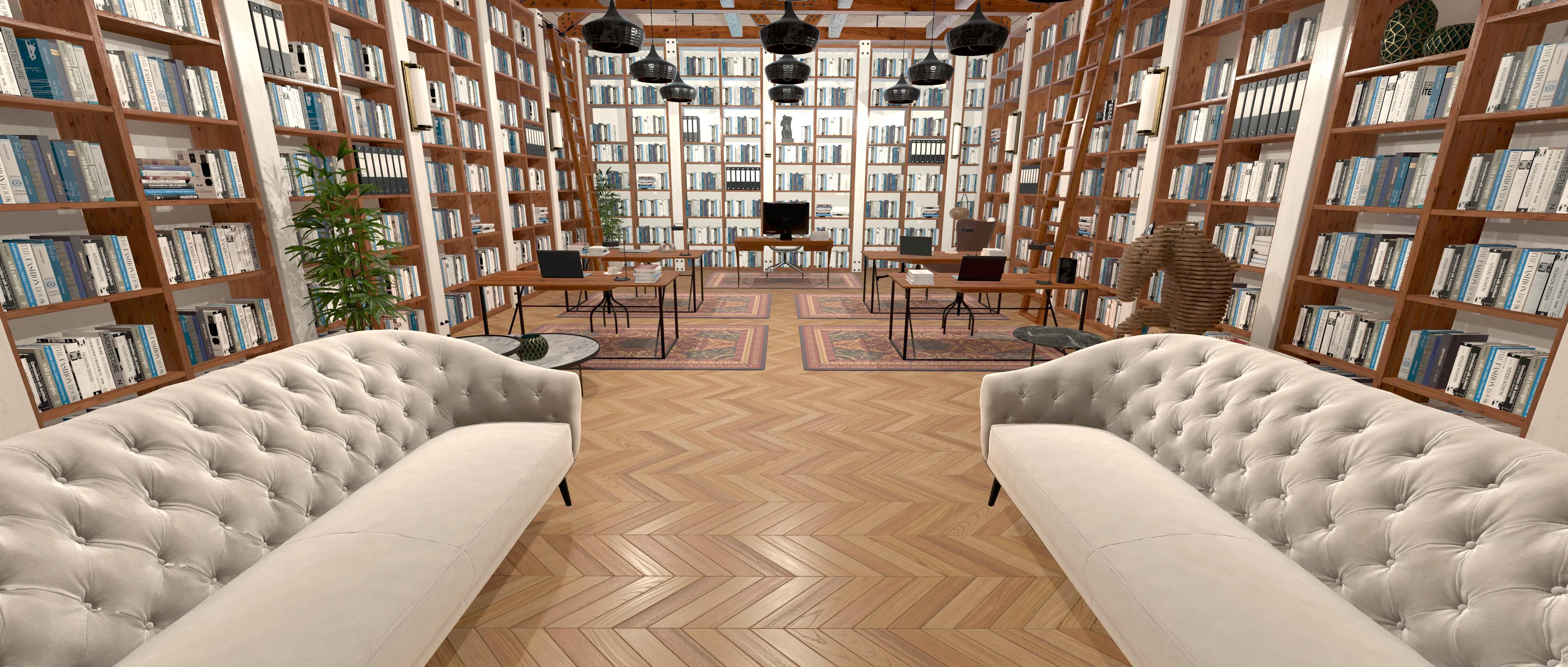}}\quad
\subfloat[Indoor1 semi-cluttered]{\includegraphics[width=\linewidth]{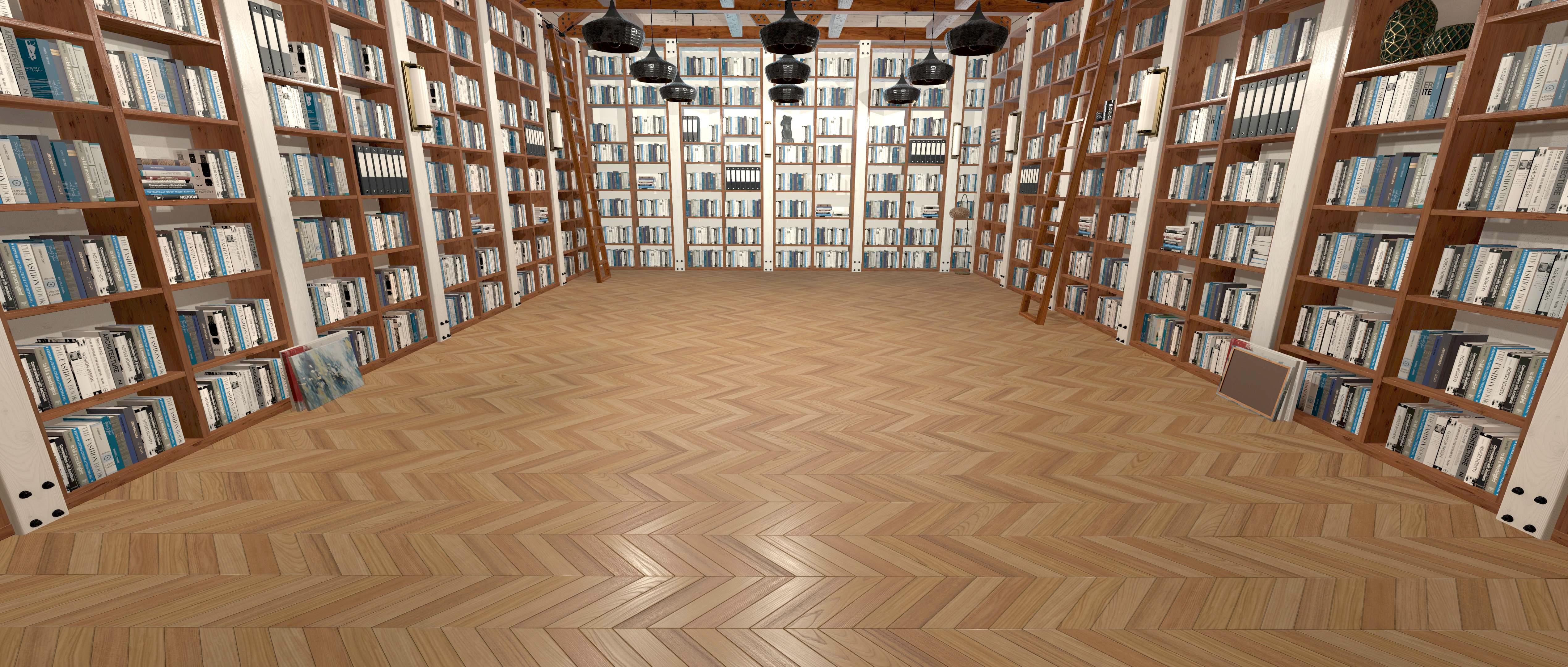}}\quad
\subfloat[Indoor1 uncluttered]{\includegraphics[width=\linewidth]{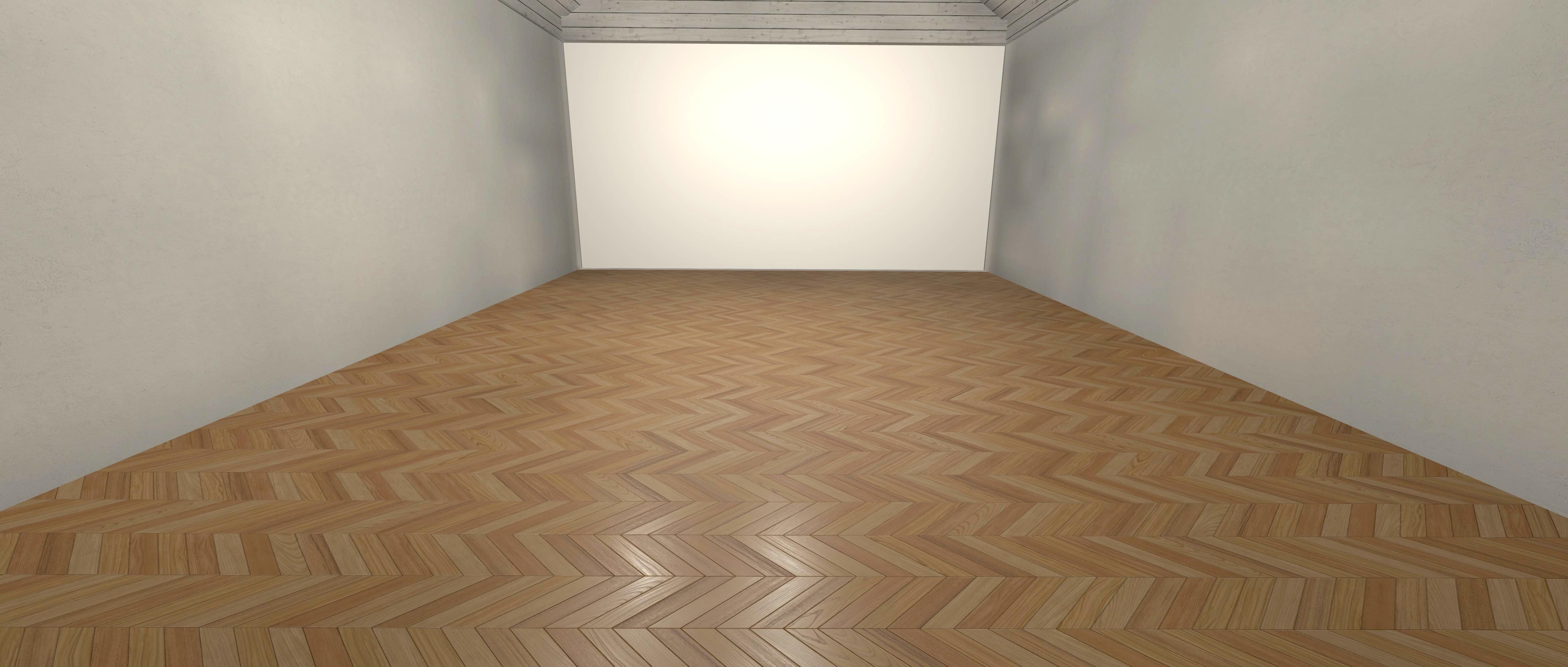}}
\caption{Indoor1 Environment Levels Of Clutter}
\label{fig:Indoor1_all}
\end{figure}

\begin{figure}
\centering
\subfloat[Indoor2 cluttered]{\includegraphics[width=\linewidth]{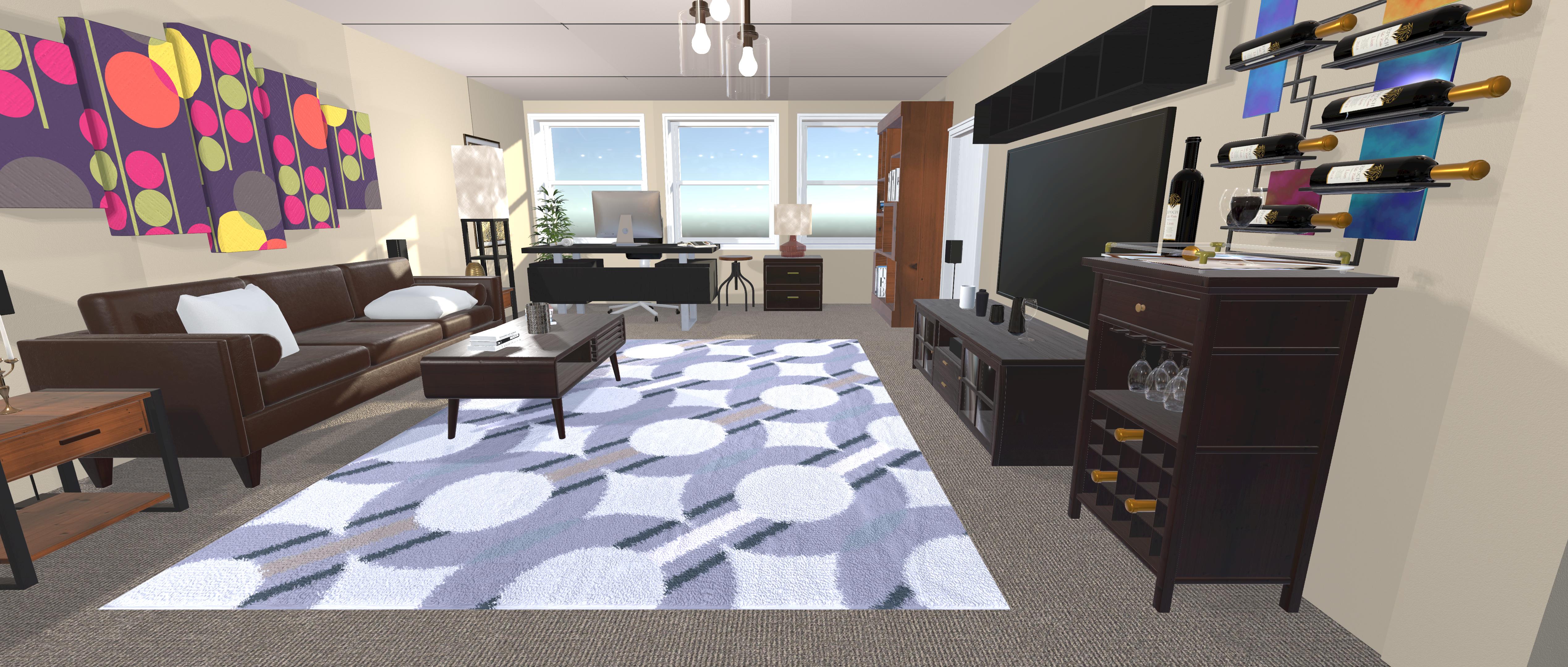}}\quad
\subfloat[Indoor2 semi-cluttered]{\includegraphics[width=\linewidth]{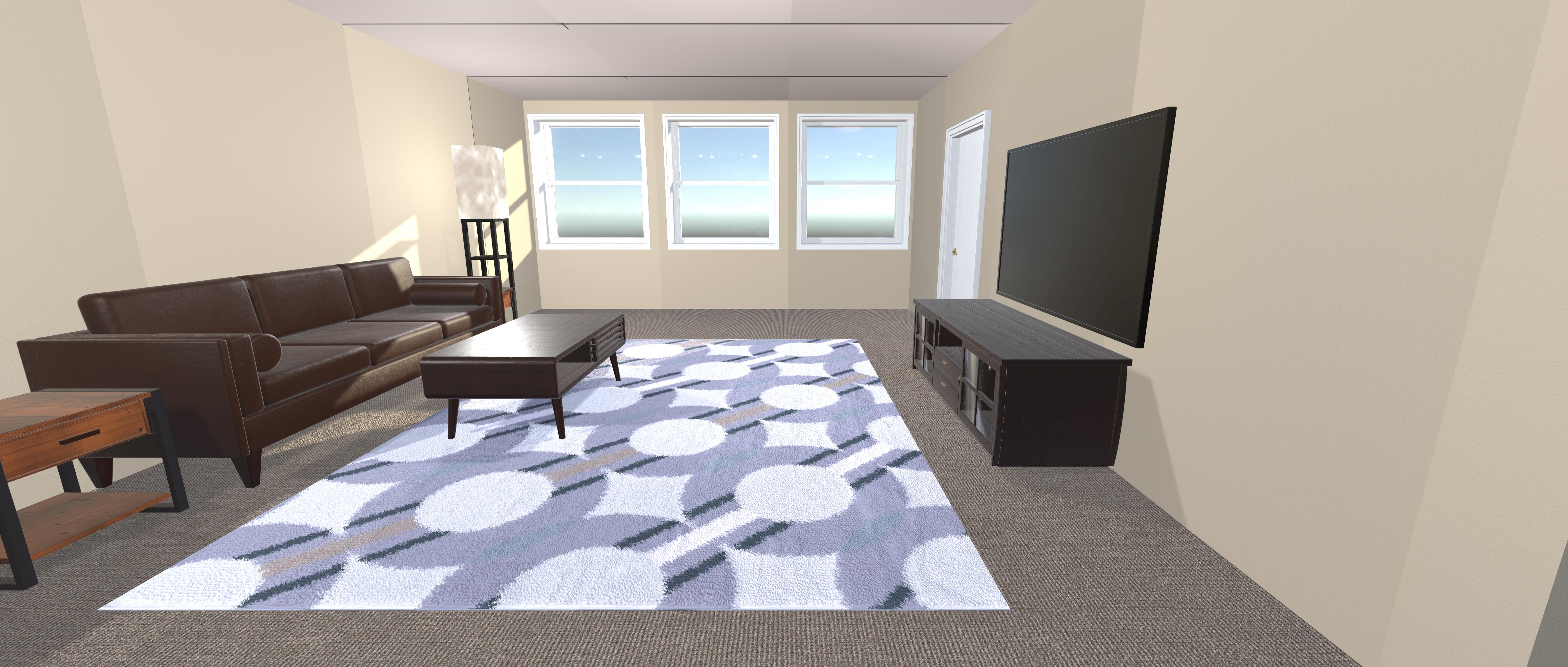}}\quad
\subfloat[Indoor2 uncluttered]{\includegraphics[width=\linewidth]{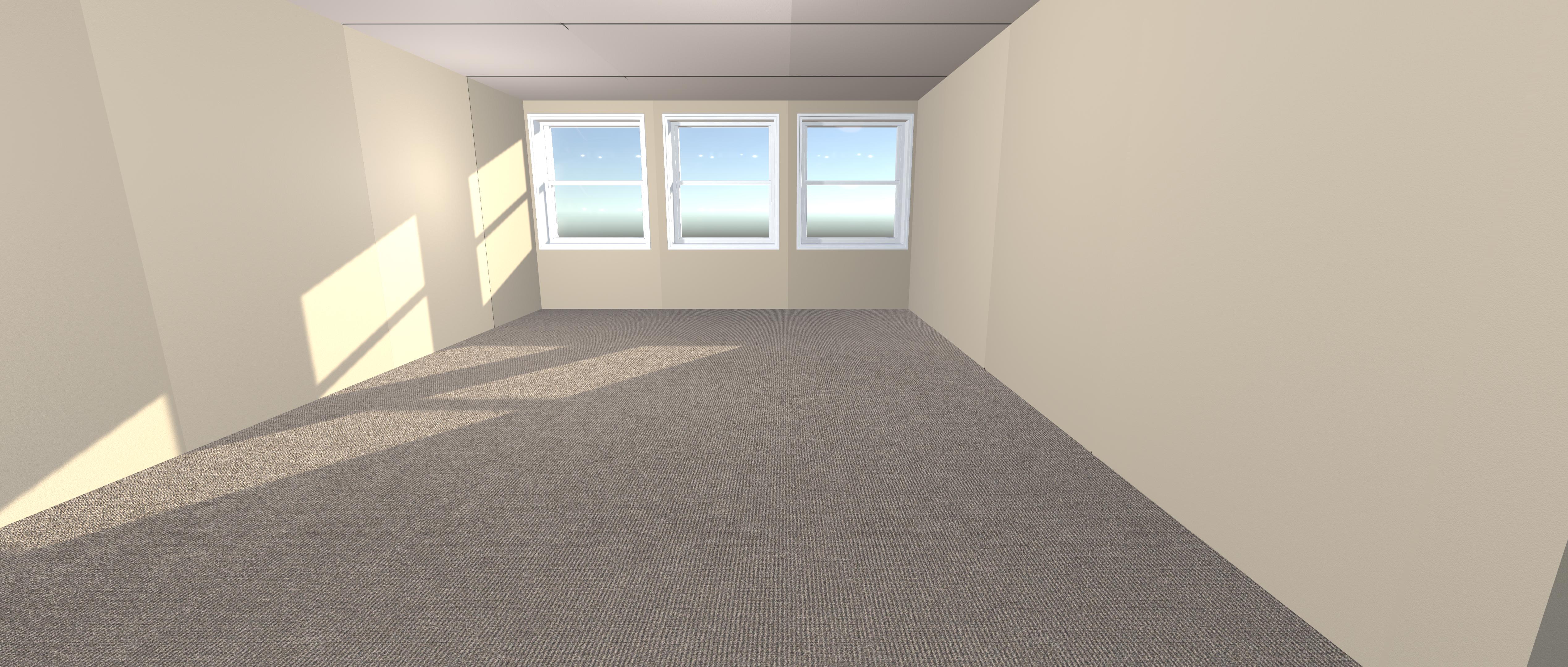}}
\caption{Indoor2 Environment Levels Of Clutter}
\label{fig:Indoor2_all}
\end{figure}

\begin{figure}
\centering
\subfloat[Outdoor1 cluttered]{\includegraphics[width=\linewidth]{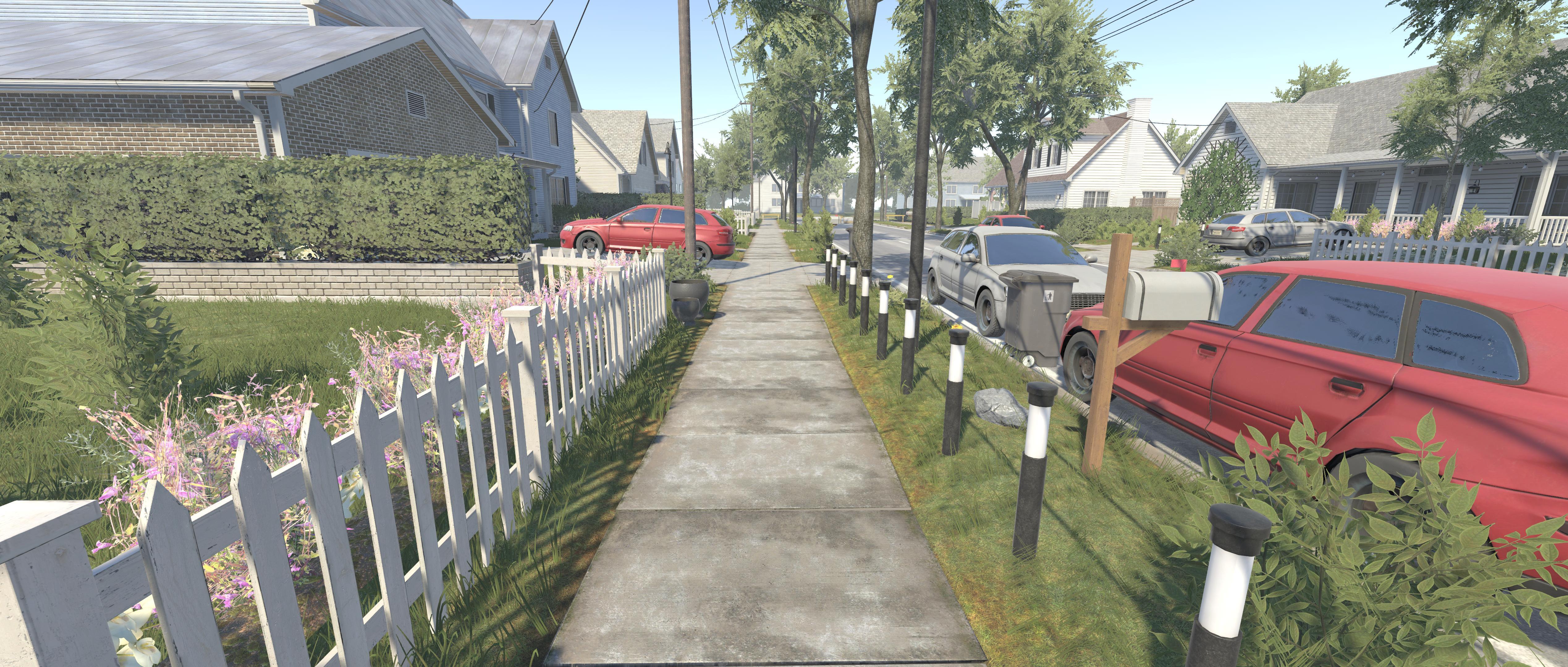}}\quad
\subfloat[Outdoor1 semi-cluttered]{\includegraphics[width=\linewidth]{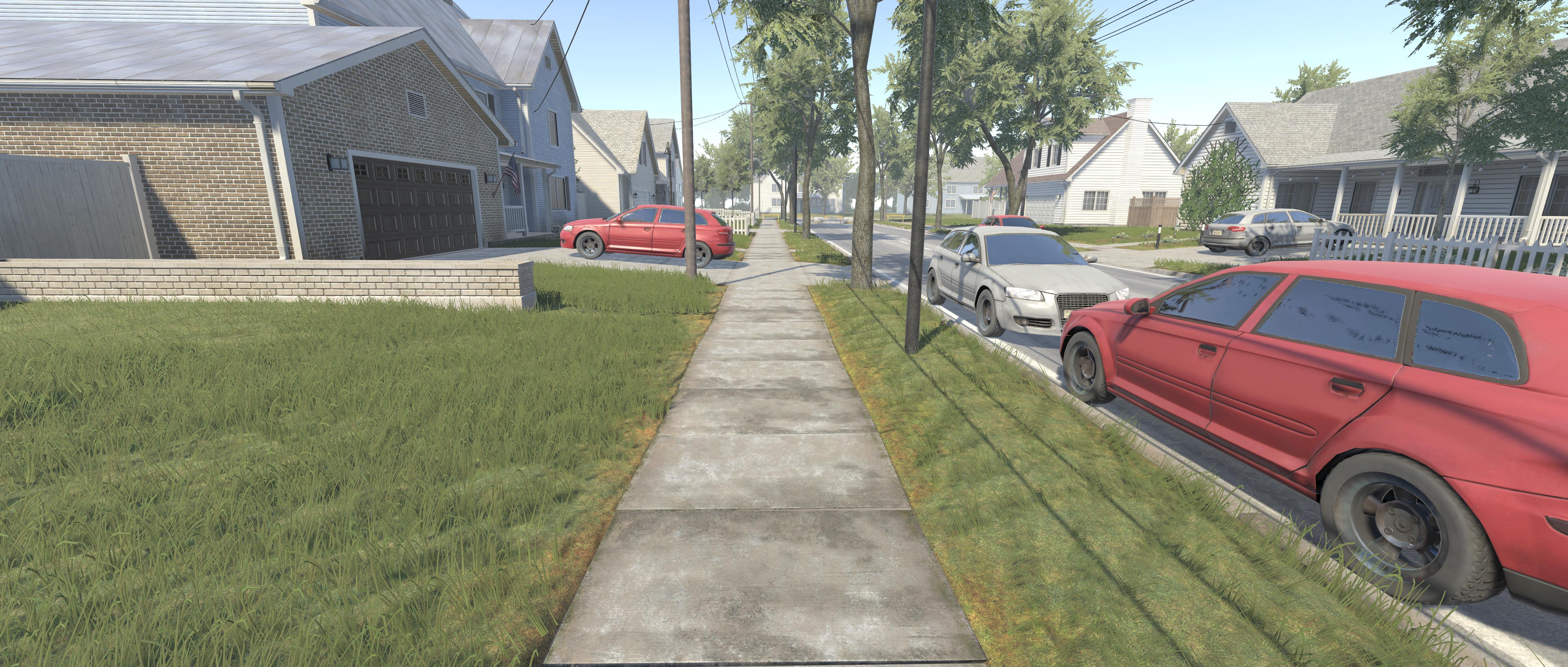}}\quad
\subfloat[Outdoor1 uncluttered]{\includegraphics[width=\linewidth]{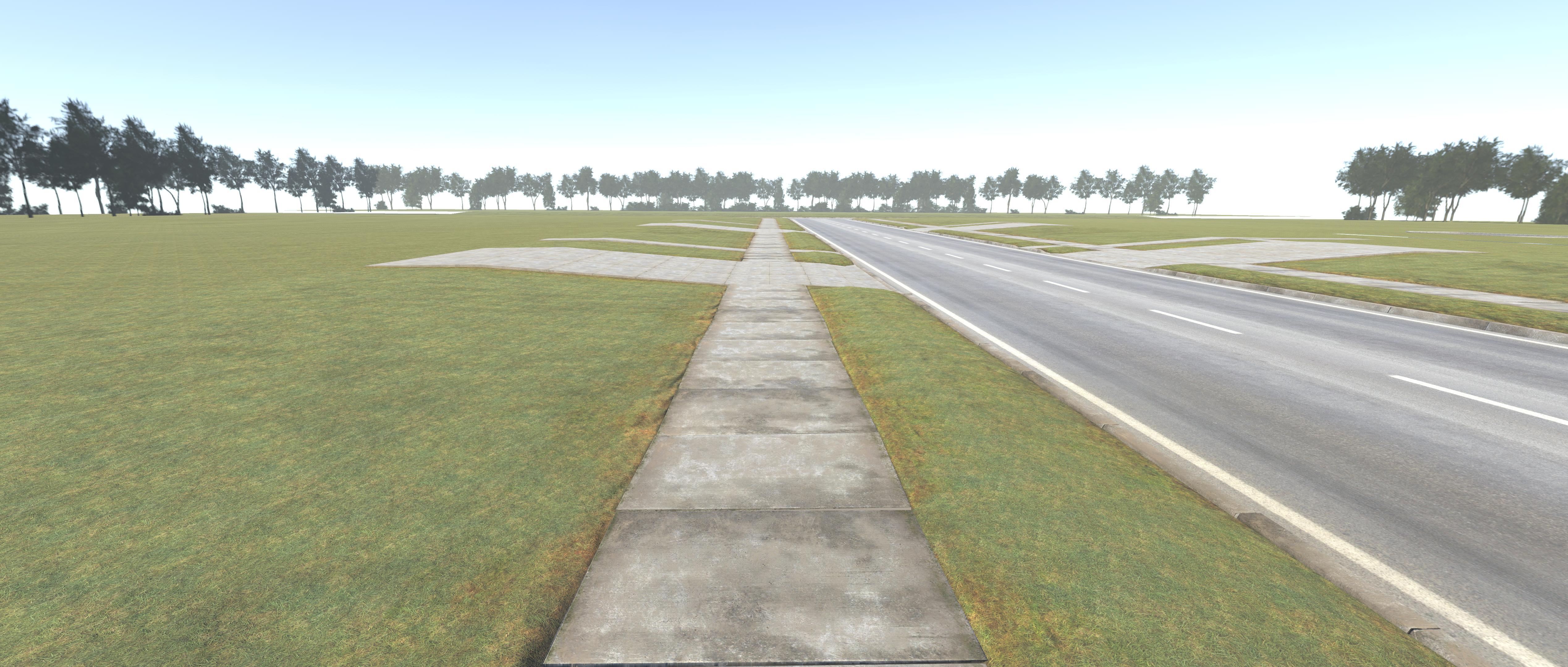}}
\caption{Outdoor1 Environment Levels Of Clutter}
\label{fig:Outdoor1_all}
\end{figure}

\begin{figure}
\centering
\subfloat[Outdoor2 cluttered]{\includegraphics[width=\linewidth]{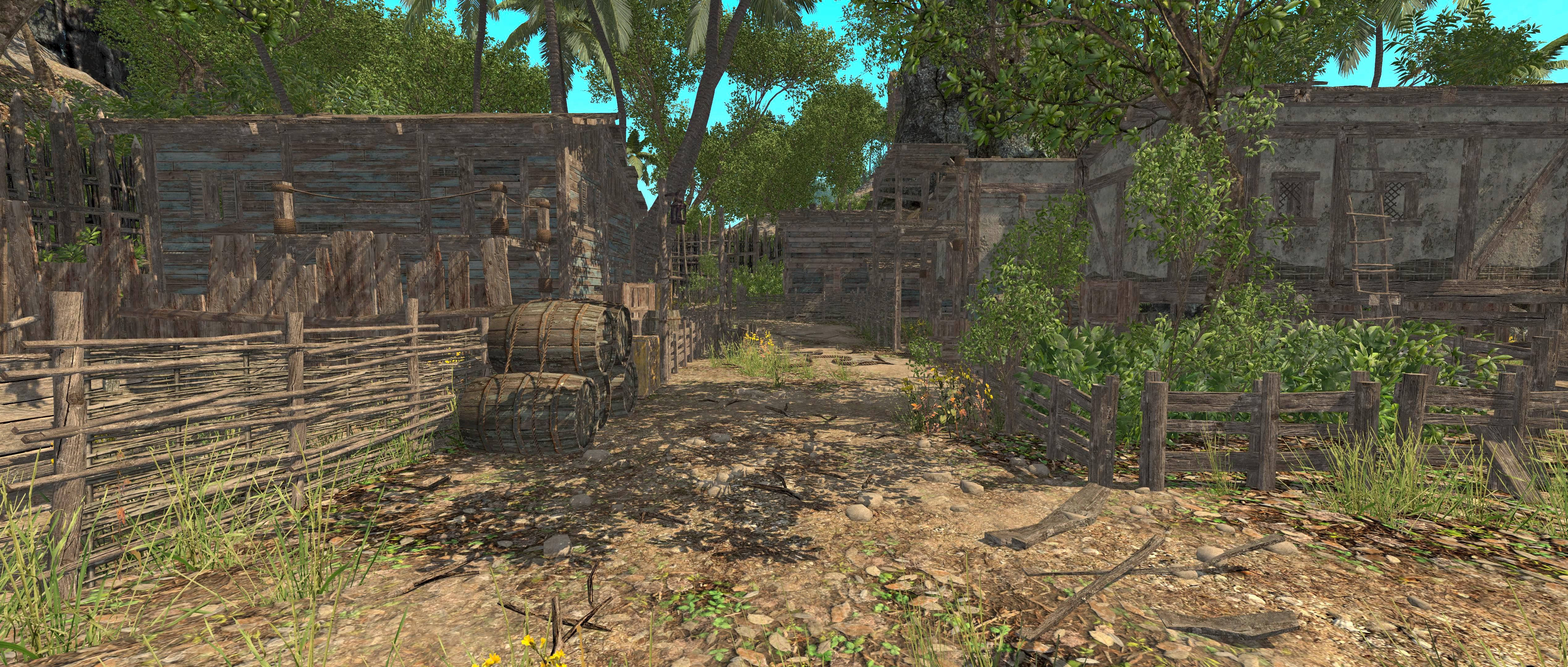}}\quad
\subfloat[Outdoor2 semi-cluttered]{\includegraphics[width=\linewidth]{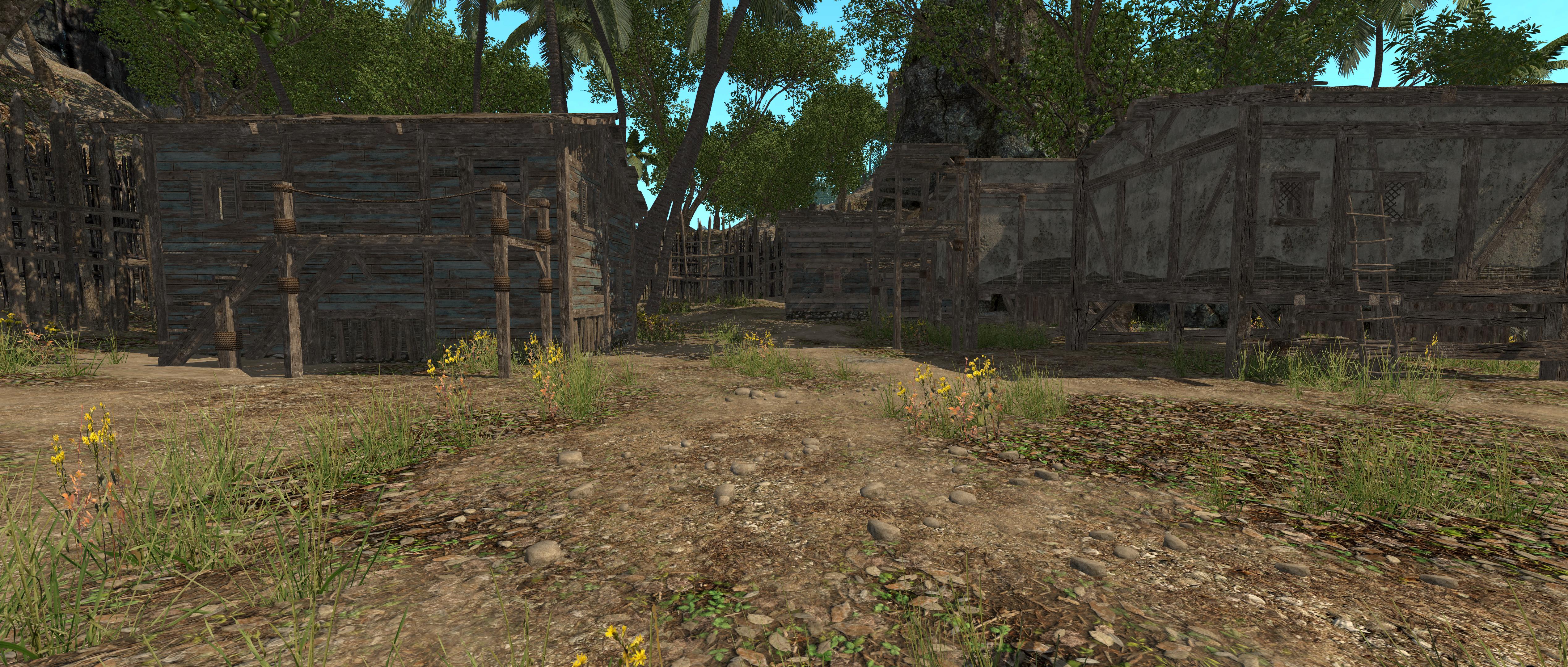}}\quad
\subfloat[Outdoor2 uncluttered]{\includegraphics[width=\linewidth]{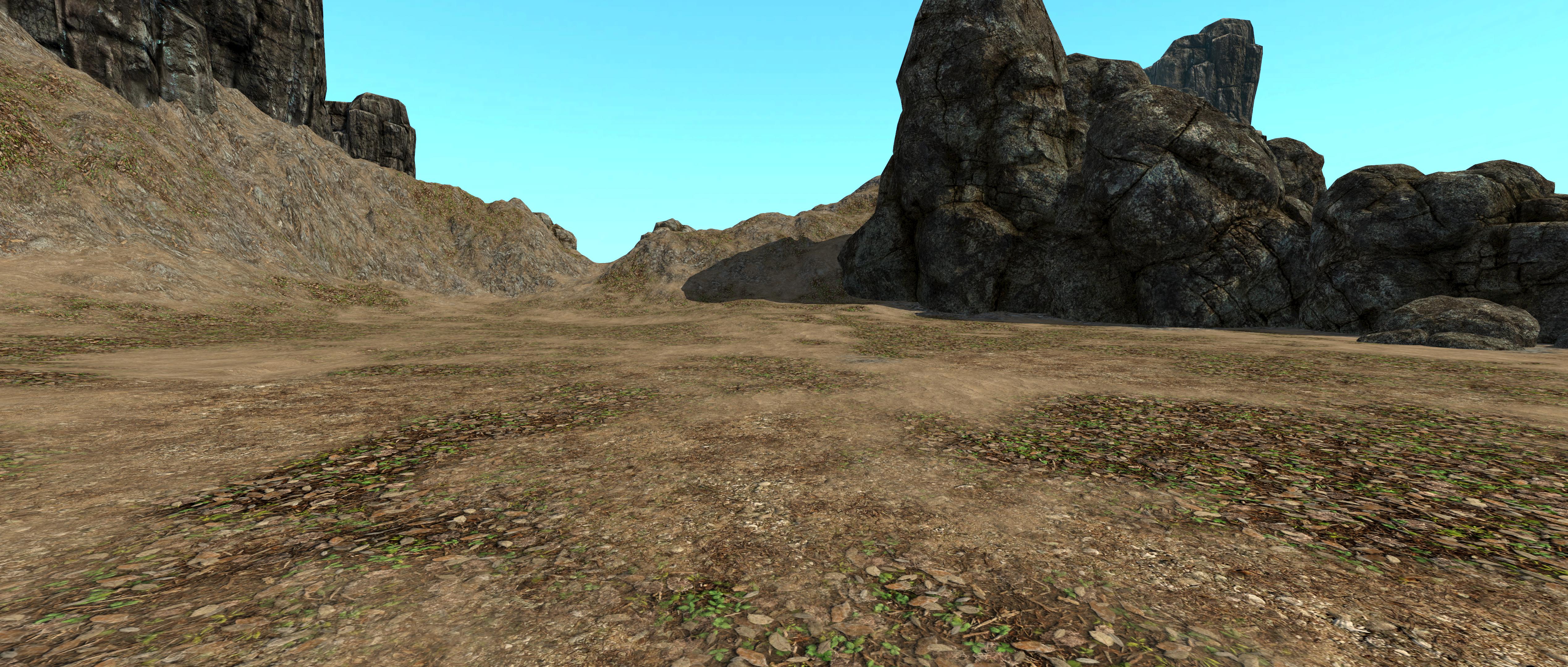}}
\caption{Outdoor2 Environment Levels Of Clutter}
\label{fig:Outdoor2_all}
\end{figure}

Four different environments were used in the study. However, each user participating in the study saw only one environment from the available ones mentioned. 
These environments were sourced from the Unity3D Asset Store\footnote{https://assetstore.unity.com/packages/3d/environments/urban/library-interior-archviz-160154; retrieved 2022-06-12}\footnote{https://assetstore.unity.com/packages/3d/environments/urban/suburb-neighborhood-house-pack-modular-72712; retrieved 2022-06-12}\footnote{https://assetstore.unity.com/packages/3d/props/apartment-kit-124055; retrieved 2022-06-12}\footnote{https://assetstore.unity.com/packages/3d/environments/pirates-island-14706; retrieved 2022-06-12} and were modified to meet our needs. 
A round-robin order was used to assign an environment to each user. The environments available consisted of Indoor1, Indoor2, Outdoor1, and Outdoor2, where every environment was designed to be realistic and was displayed to the participants using Unity3D. 
The \textit{Indoor1} VE (see Figure~\ref{fig:Indoor1_all}) was a library with a 10m$\times$7m size, a ceiling with a 4m height, along with desks, sofas and bookshelves along three sides of the room. 
The \textit{Indoor2} VE (see Figure~\ref{fig:Indoor2_all}) was a 10m$\times$5m sized living room of an apartment and had a ceiling of a 3m height along with windows in a side of the room, a carpet, and furniture. 
The \textit{Outdoor1} VE (see Figure~\ref{fig:Outdoor1_all}) was a sidewalk in a suburban neighborhood. This environment was in daylight and it contained a few cars parked on the street and on one side a fence. 
The \textit{Outdoor2} VE (see Figure~\ref{fig:Outdoor2_all}) was an island in daylight that contained trees, ropes, and barrels along with cottages and boxes made out of wood. The items described for each environment were positioned in a way so that they did not interfere with the walking area.

\hfill

\hfill
\subsubsection{ Within-Subjects Variables}
\begin{figure}
  \centering
  \includegraphics[width=.95\linewidth]{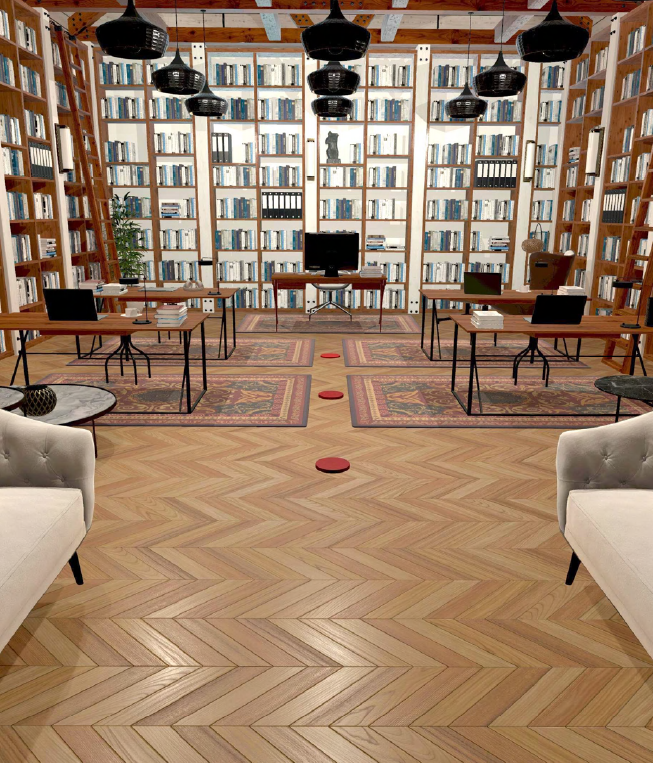}
  \Description{All Three Targets Positions Possible In The Study}
\caption{All Three Targets Positions Possible In The Study Seen From a User's Point Of View}
\label{fig:Target_all}
\end{figure}

We designed three clutter levels for each environment. The clutter levels consisted of: \textit{1:uncluttered, 2:semi-cluttered, and 3:cluttered}. 
We defined \textit{clutter} as the number of objects visible to the user at each scene.
Three different FOVs were simulated inside the same HMD using the software. 
The chosen FOV levels were picked based on the physical FOVs of popular VR headsets:
Pimax 5K (165° $\times$ 110°), HTC Vive/Oculus Quest (110° $\times$ 110°), and nVisor ST60 (45° $\times$ 35°).
The target object chosen for the study was a red cylinder, which had a 10cm diameter and a 5cm height. 
The target was on the ground and the distance of the target from the user's starting point was either 3m, 4.5m, or 6m. The target design was similar to a previous study done by Masnadi et. al~\cite{masnadi2022effects}. Figure~\ref{fig:Target_all} demonstrates all of the possible target distances in an environment. The choice of these distance ranges was made based on prior blind-walking user studies~\cite{sahm2005throwing,masnadi2022effects}. To prevent tiredness and reduce the learning effect along with reducing the number of trials, we chose three target distances instead of four (3m, 4m, 5m, and 6m) while keeping the range unchanged.
To add more realism to the study, we did set the target object to cast and receive shadows with the aim of offering the participants depth cues that are realistic in addition to making the target blend in with the environment.
In addition, the target choice was also made with the purpose of enabling the participants to even walk over it or get near it without being afraid of encountering an obstacle. 

We used a rectangular boundary for each scene as the \textit{safe area}, such that the safe area refers to the area in the VE where the user can be placed along with the target without the user encountering a collision with the other virtual objects present in the scene while walking to the target. We designed a randomization of the user's location in each of the environments at each trial. In every trial, the camera, which represents the starting point of the user, had its transform parameters randomized such that both the targets set and the start point would both be within the described safe area. Furthermore, the FOV and the target placement were different for every two consecutive trials. Randomizing the starting point was done to reduce the possibility that the user memorizes the number of steps. This randomization also ensures environmental variations that promote the reduction of environment-based effects.

The combination of the factors--consisting of the clutter level, FOV, and target distance-- resulted in 27 different conditions. Every condition was presented three times to every participant which resulted in 81 blind-walk trials, where the order was randomized for every participant, such that no consecutive trial had a common FOV or target distance in order to minimize the memorization effect. During each measurement, the error, which is the distance of the user's position from the actual target, was recorded. A positive error value was recorded if the user walked passed the target, and a negative error value was recorded if the user walked short of the target.


\begin{figure}
  \centering
  \includegraphics[width=.95\linewidth]{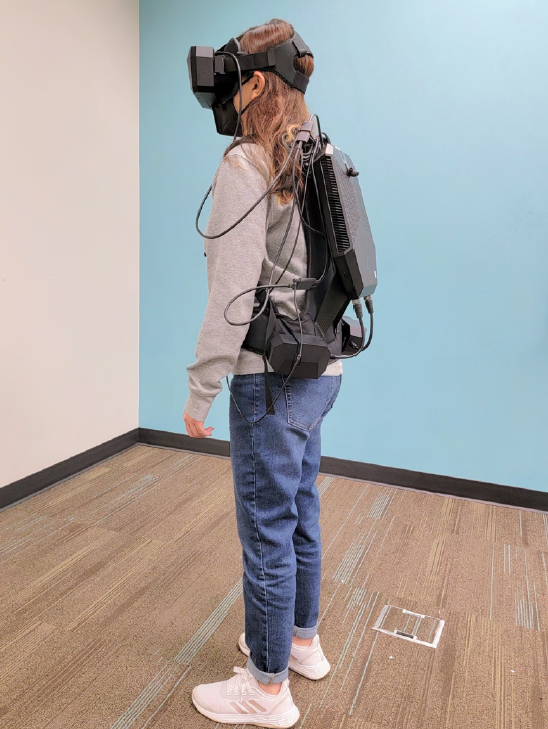}
  \captionof{figure}{A user wearing the backpack and the HMD}
  \label{fig:user}
\end{figure}


\subsection{Participants}
The participants were recruited from the university population and their number totaled 60 users (46 male, 13 female, and one non-binary) with ages ranging from 18 to 34 (M=21.45, SD=3.44) along with heights ranging from 152cm to 208cm (M=174.12, SD=11.55). Among the users, 35 were wearing glasses or contact lenses and kept them on during the study. Furthermore, participants were asked to share their VR experience in a scale ranging from 1 (least experienced) to 5 (most experienced), which resulted in M=2.57 and SD=1.23.

\subsection{Apparatus}
The study was performed using a Pimax 5K Plus VR headset, which has a 165° (horizontal)$\times$110° (vertical) FOV and a resolution of $2560 \times 1440$ pixels per eye. In addition, the stated headset has a 120hz refresh rate and a weight of 470g. We used an HP Z VR Backpack equipped with an Intel 7820HQ CPU, an NVIDIA Quadro P5200 GPU, and 32GB memory. Moreover, we added a tiny speaker on the backpack near the back of the user's head with the aim of providing verbal instructions. Using the backpack required adding external batteries to it, which result at the end in a weight of 4.35kg -- including the weight of the backpack itself, batteries, and the harness.
We used a canvas through the Unity3D editor in order to limit the user's FOV, which was set at a distance of 1cm from the cameras. We opted for using cutouts in a black plane to recreate the desired FOVs.


We designated an empty area in our closed laboratory to perform this study. The room dimensions were $6m(w)\times9m(l)\times3.3m(h)$ and the empty area in the middle of the room was $4m(w)\times 9m(l)$. The furthest target was 6m away from the user and there was a 2m distance between this target and the closest non-study object. In addition, the chosen Apparatus relies on the use of SteamVR\footnote{https://partner.steamgames.com/doc/features/steamvr/info}, which shows a visual safeguard 
when the user gets near the borders of the study area. This Safeguard was kept enabled as a precaution, yet based on the empty area we chose to perform the study, the Safeguard never appeared and consequently did not influence the perception of distance. 



To avoid interfering with the visual display perceived by the user and prevent a negative impact on the performance, we have designed a remote based user monitoring tool for the user study.
We designed a system to monitor the behaviour of the participant and visualize what they are viewing in real-time on another computer. This system transfers compressed images from the headset to the monitoring computer through network without a negative impact on performance nor the ongoing study in general. Along with transmitting images, the designed tool transmitted trial information, which helped the researcher have a better grasp of the ongoing user activity. The study code was made public on Github \footnote{Anonymous}. The posted project provides a Unity prefab, which allows performing the study by adding it to any Unity 3D scene.

\subsection{Procedure}
After receiving the user's consent to participate in the study, we proceeded to evaluate their visual acuity to ensure that they were eligible to participate in the study using a Snellen chart (see \cite{sue2007test}). A score better than 20/32 for each eye was required to participate. This was necessary to be able to distinguish visual details in the scene. Moreover, users wearing glasses or lenses were asked to keep them during the study. All the participants that failed the vision evaluation were dismissed. On the other hand, the participants that had an adequate vision score were asked to complete a survey that asked for the following information: age, height, gender, handedness, and experience with VR.

After these steps were completed, the investigator proceeded to explain the study-related tasks in detail and the purpose of each step and requirement in the study. In addition, the participants were shown the gear to be used and how to adjust it. Before wearing the backpack and headset, users were asked if they had any questions so the researcher could clarify any present inquiry.

Users were prompted to wear the backpack and adjust it as needed. Then, we gave them the headset and helped them wear it and adjust it so it was comfortable and well-placed on their head. Figure~\ref{fig:user} shows a user with the HMD used in the study along with the HP backpack. After the user was equipped with the gear and was ready, we informed them that we would turn off the lights so they would not have any indication from the outside that could help them in the target distance evaluation, then we started the evaluation process and initiated the data collection process.

At first, the user was asked to begin from a specific start position, which initially was logged to be the starting position and they had to return to the start position to begin the next trial.
The user looked at a black screen at this point. Then, whenever the user was ready, they let the researcher know by saying the word ``ready'' and then we showed them the VE such that the FOV and target position were automatically changed following a pre-processed trial sequence. For five to six seconds the user visualized the environment and tried their best to understand the distance between them and the target. When they were ready to walk, they informed the researcher by saying ``ready''. Afterward, the researcher initiated the walking process by hiding the environment through a button click on a wireless keyboard that made the environment invisible. Simultaneously, a clear and faint audio source was played from the backpack speaker that says ``go''. Then, the participant started walking with closed eyes towards the target and they stopped as soon as they realized the current location was the one where the perceived target was. While the participant was walking, all the content of the VE was hidden and a black screen was displayed to ensure that in the case the participant opens their eyes, no information related to the environment would be given while walking to the target. When they reached their desired spot for the current trial, the participant said ``here'' to confirm verbally with the researcher that they reached the target. The researcher logged the position through a button click on the wireless keyboard. Simultaneously, a faint audio source is played that said ``done'' to inform the participant that they can open their eyes and follow a red arrow that showed up near their feet to guide them back to the starting position. This guidance arrow was always attached to the participant's feet, and it was always pointing to the start position. When the arrow showed up, the participant followed its direction until they see a green-colored arrow that marks the starting position. This green arrow is an alignment arrow that helps the user correctly align with the real-world room. The user moved so that the red (direction arrow) and green (alignment arrow) arrows were overlapping and in the same direction. When the alignment was correct, only the green arrow remains and then the user informed the researcher that they were ready to view the next scene by saying ``ready''. After getting the confirmation, the researcher pressed a button key on the wireless keyboard to show the VE leading to the start of a new trial.

Since this user study was performed during COVID-19, we prioritized the safety of each one of the users and researchers. Therefore, we relied on the previously stated arrows to keep a safe distance between the researcher and users along with reducing their physical interaction. Through the arrows used, the researcher did not need to give any indication to the users about how to navigate back to their starting position nor to manually intervene to bring them back to it. Moreover, the study design, as explained, reduced the chance of having verbal interactions between the researcher and participants, which prevented the user from receiving audio cues from the outside. In addition, the equipment used in the study was sanitized and cleaned before and after each user study, while also both the researchers and users were required to wear masks.


Every user evaluation was allocated a one-hour time slot and the user received \$10 USD in cash at the completion of the study.

\subsection{Hypotheses}
For the last couple of decades, researchers have not concluded that the FOV is a crucial factor that brings about the underestimation of distance in VR \cite{willemsen2009effects, renner2013review}. Nevertheless, new research has presented that modern technology eradicates distance underestimation in VR, which can correlate to improving the FOV \cite{kelly2017perceivedSpace,kelly2022distance}. Furthermore, other works have found that distance underestimation is negatively impacted by reduced FOVs \cite{pfeil2021distance, masnadi2021field}. Moreover, it is important to mention that the perception of distance is also influenced by other factors such as the environment itself. Creem-Regehr et al. and Masnadi et. al came to the finding that distance perception is more precise in indoor settings in contrast to outdoor environments \cite{creem2015egocentric, masnadi2022effects}. Adding more objects into an environment is a practice that is shared by several designers, as they intend to improve distance judgment in their environments. However, there is no evidence that adding more clutter to environments provides visual cues that improve distance perception~\cite{pfeil2020gender, paraiso2017entourage}. As a consequence, we designed and conducted our study based on the described parameters and our hypotheses are:

\begin{itemize}
    \item[H1:] Participants will more accurately estimate distances in more cluttered environments.
    \item[H2:] Participants will more accurately estimate distances when viewing indoor environments.
    \item[H3:] Participants will more accurately estimate distances with wider FOVs.
    \item[H4:] Clutter and FOV have independent effect on distances judgement.
\end{itemize}

\section{Results}
In this section, we will show the results of our study along with the data analysis using ANOVA. The errors are reported in cm.
\subsection{Repeated Measures ANOVA Results:}
We performed a Shapiro-Wilks normality test on the data which showed that the
data were not normally distributed (p<.01). 
Therefore, we used the ARTool to transform the gathered data~\cite{wobbrock2011aligned}. Afterward, we performed a mixed ANOVA test which had the environment as the between-subject variable and clutter, FOV, and distance as the within-subject variables.
We found a main effect of clutter, FOV, distance, and environment. An interaction effect between Clutter$\times$Environment, FOV$\times$Distance and Distance$\times$Environment was found. However, we did not find an interaction effect between Clutter$\times$FOV. Table \ref{table:MainEffectsAndInteractionEffects} shows the results of the omnibus test in detail.

\begin{table*}
\centering
	\caption{Repeated Measures ANOVA results.}
	\begin{tabular}{l|lll}
		\textbf{Effect on Error} & \textbf{ANOVA Result}\\
		\midrule
		\midrule
		\multicolumn{4}{c}{Main Effects} \\
		\midrule
        Clutter  &      $F(2,59) = 4.673,$&     $p = .009,$&    $\eta^2_p = .026$ \\
		FOV  &          $F(2,59) = 29.748,$&    $p < .001,$&    $\eta^2_p = .145$ \\
		Distance  &     $F(2,59) = 254.180,$&   $p < .001,$&    $\eta^2_p = .591$ \\
		Environment  &  $F(3,56) = 17.598,$&    $p < .001,$&    $\eta^2_p = .231$ \\
		\midrule
		\midrule    
		\multicolumn{4}{c}{Interaction Effects}\\
		\midrule
		Clutter$\times$FOV  &           $F(4,173) = 2.519,$&$p = .051,$&$\eta^2_p = .013$ \\
		Clutter$\times$Distance  &      $F(4,173) = 1.152,$&$p = .331,$&$\eta^2_p = .007$ \\
		Clutter$\times$Environment  &   $F(6,352) = 2.557,$&$p = .019,$&$\eta^2_p = .042$ \\
		FOV$\times$Distance &           $F(4,173) = 3.434,$&$p = .009,$&$\eta^2_p = .019$ \\
		FOV$\times$Environment &        $F(6,352) = 1.399,$&$p = .214,$&$\eta^2_p = .023$ \\
		Distance$\times$Environment  &  $F(6,352) = 14.935,$&$p < .001,$&$\eta^2_p = .203$ \\
		\midrule
		Clutter$\times$FOV$\times$Distance  & $F(8,169) = .238,$&$p = .984,$&$\eta^2_p = .001$ \\
		Clutter$\times$FOV$\times$Environment  & $F(12,525) = 1.381,$&$p = .170,$&$\eta^2_p = .023$ \\
		Clutter$\times$Distance$\times$Environment  & $F(12,525) = 1.298,$&$p = .215,$&$\eta^2_p = .000$ \\
		FOV$\times$Distance$\times$Environment  & $F(12,525) = 1.350,$&$p = .186,$&$\eta^2_p = .022$ \\
		
		\midrule
		Clutter$\times$FOV$\times$Distance$\times$Environment  & $F(24,513) = 1.445,$&$p = .076,$&$\eta^2_p = .024$ \\
		\midrule
	\end{tabular}
	\label{table:MainEffectsAndInteractionEffects}
\end{table*}

\subsection{Implication Of Findings:}
Our study resulted in acquiring insight on distance perception in multiple levels of clutter through
different FOVs. Below, we will describe in details the implications of our results.

\begin{figure}
\centering
\begin{minipage}{.48\textwidth}
  \centering
  \includegraphics[trim=0 10 0 0,width=\linewidth]{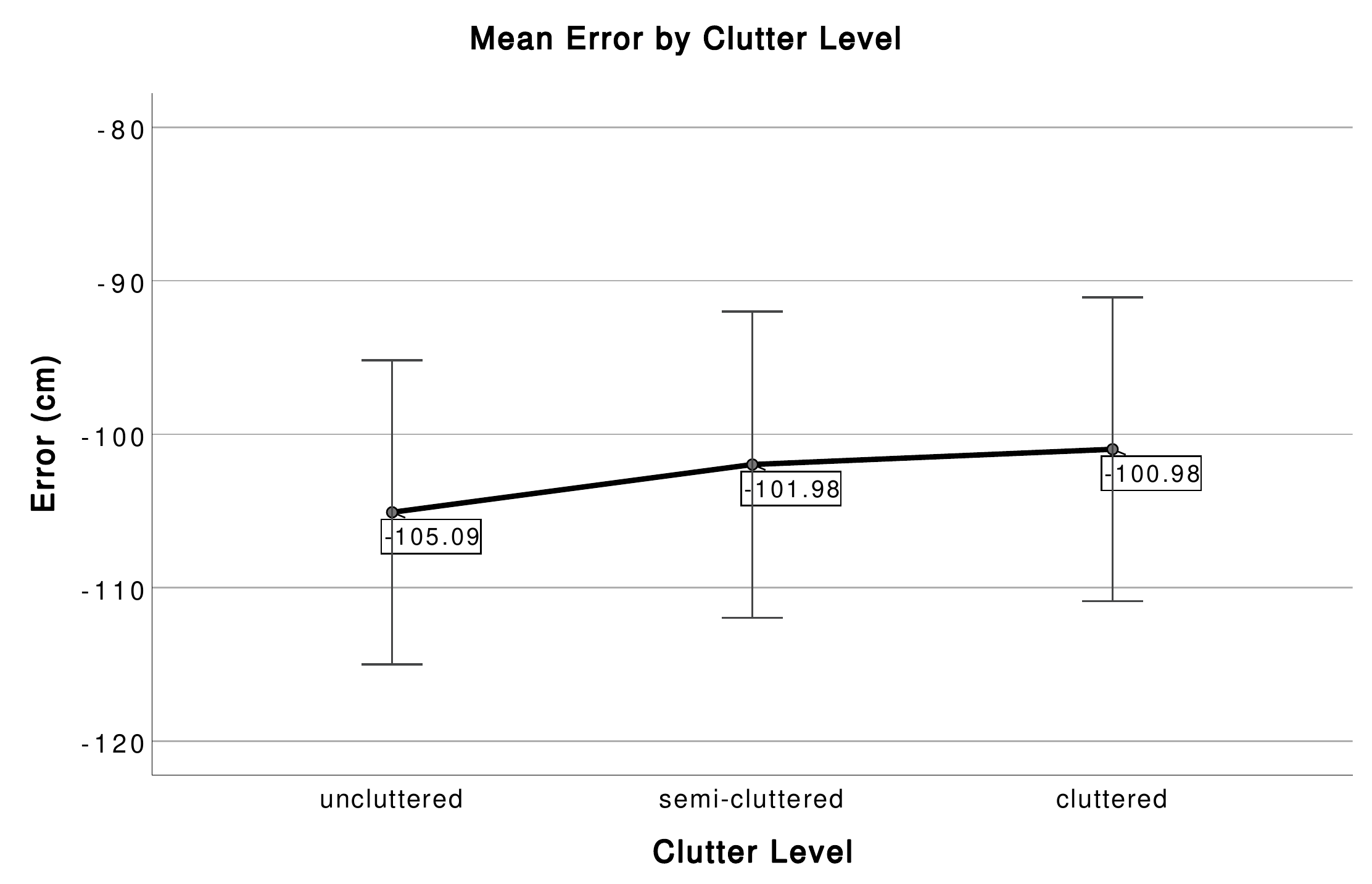}
  \captionof{figure}{The figure shows the mean error of different clutter \\levels. (95\% CI)}
  \label{fig:MEDifferentClutterLevels}
\end{minipage}%
\hspace{0.5cm}
\begin{minipage}{.48\textwidth}
  \centering
  \includegraphics[trim=0 0 0 0,width=\linewidth]{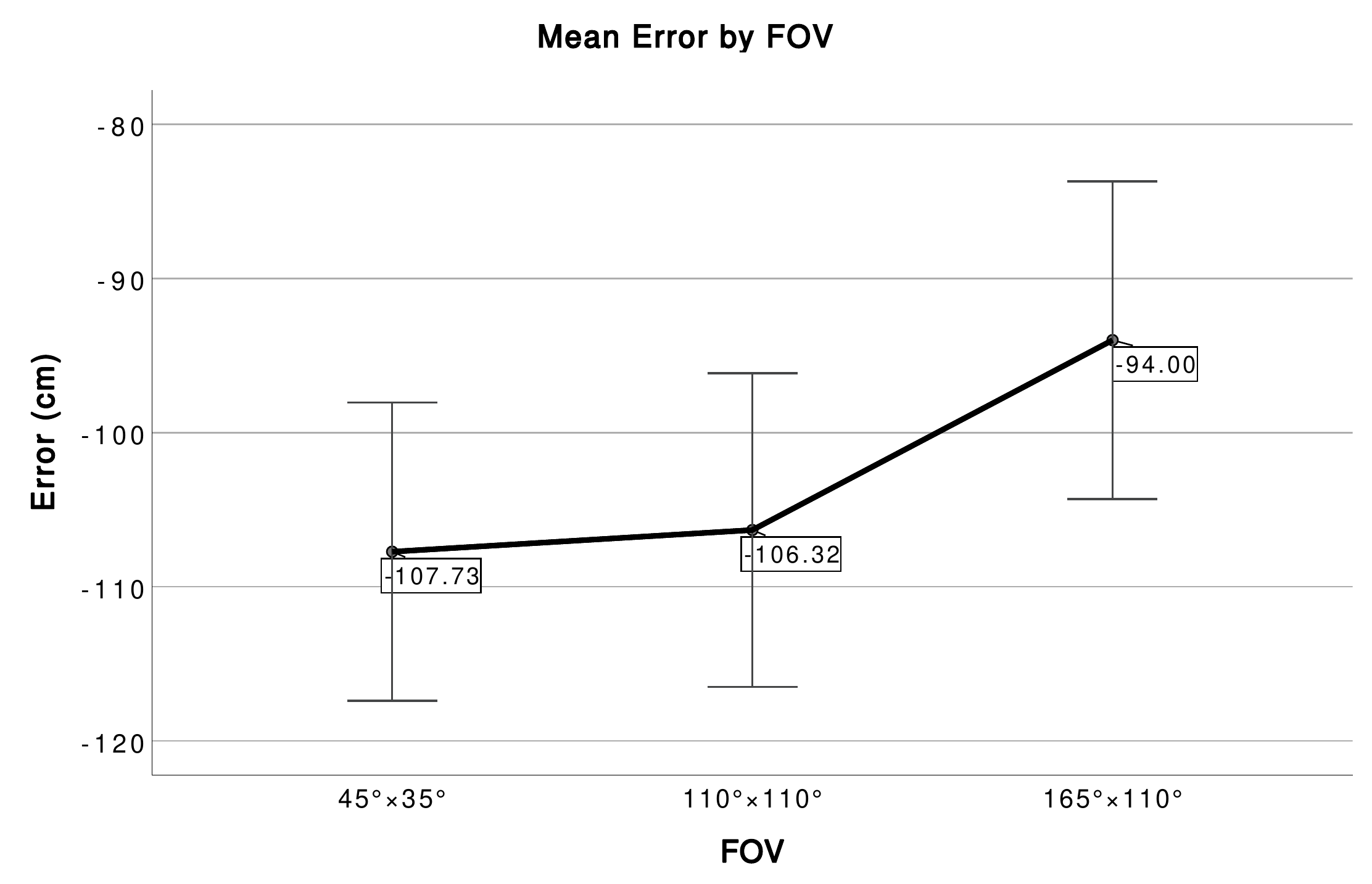}
  \captionof{figure}{The figure shows the mean error of different \\FOVs. (95\% CI)}
  \label{fig:MEDifferentFOVs}
\end{minipage}
\end{figure}





\begin{figure}
\centering
\begin{minipage}{.48\textwidth}
  \centering
  \includegraphics[width=\linewidth]{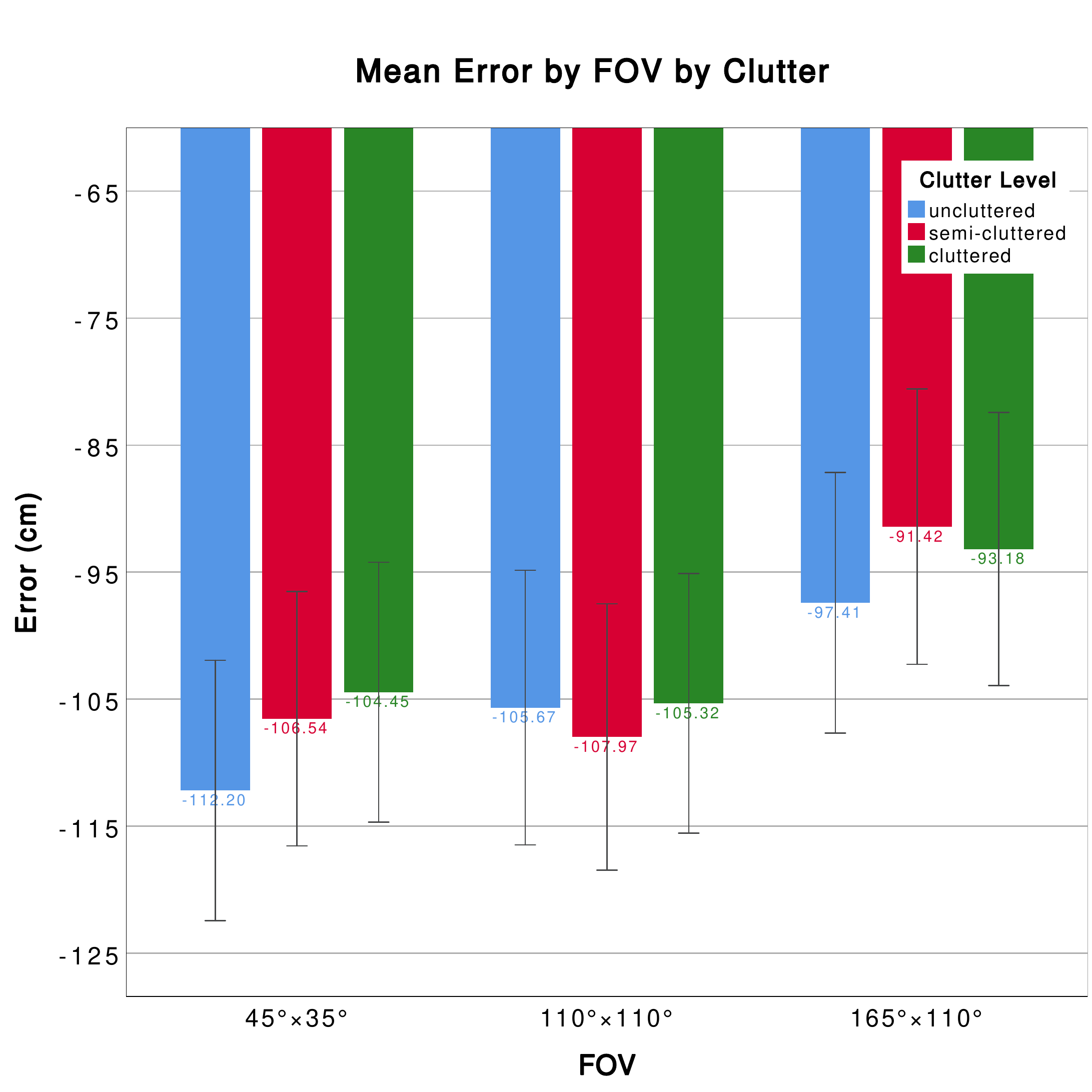}
  \captionof{figure}{The figure shows the mean error (cm) by clutter over each FOV. (95\% CI)}
  \label{fig:MEbyClutterbyFOV}
\end{minipage}%
\hspace{0.5cm}
\begin{minipage}{.48\textwidth}
  \centering
  \includegraphics[width=\linewidth]{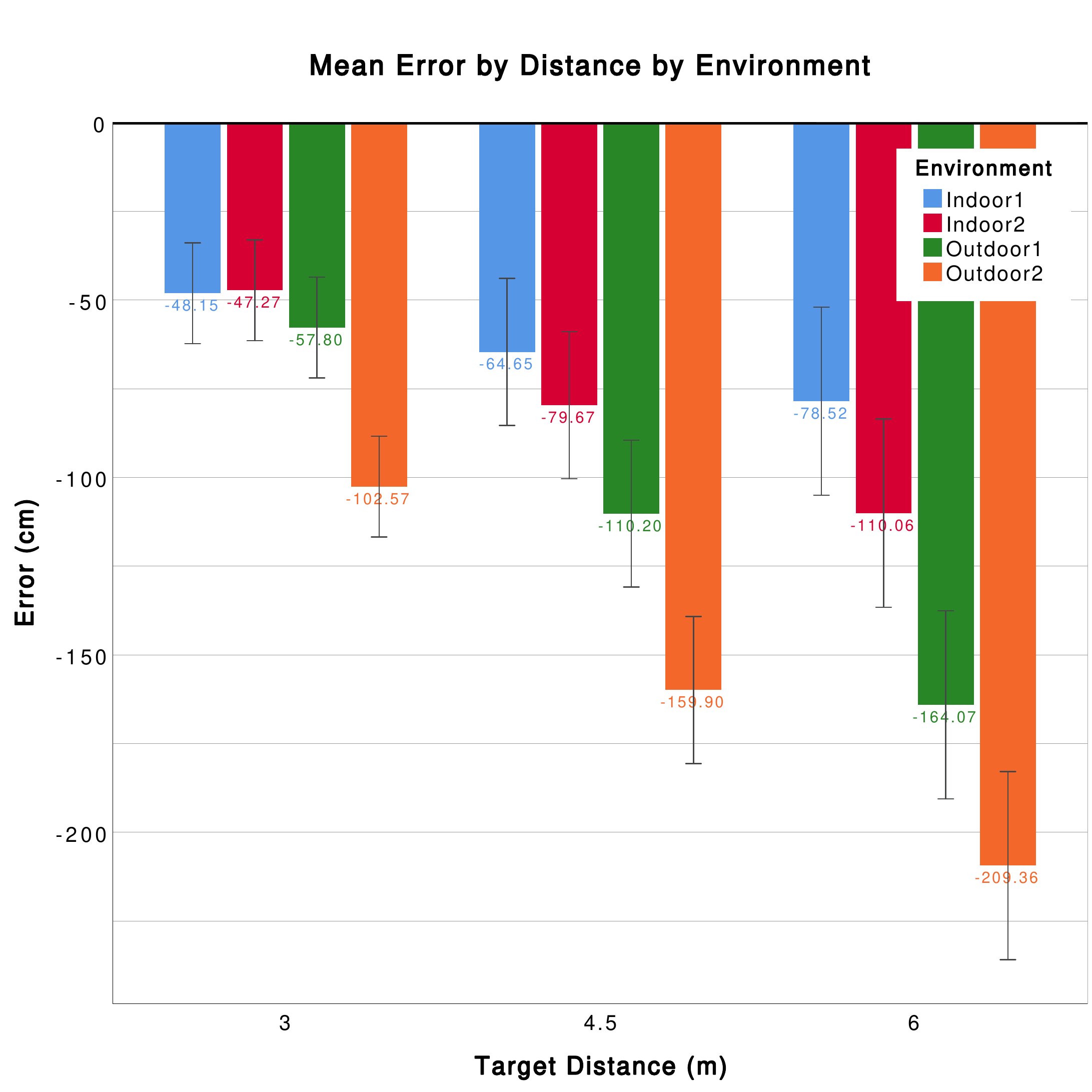}
  \captionof{figure}{The figure shows the mean error by target distance over each environment. (95\% CI)}
  \label{fig:MEbyDistancebyEnv}
\end{minipage}
\end{figure}



\subsubsection{Main Effect of Clutter}

We found a main effect of clutter on perception of distance (p = .009). Higher clutter in the environment results in more precise estimation and judgment of distances and reduces underestimation (see Figure \ref{fig:MEDifferentClutterLevels}). 
A significant difference was found between 
uncluttered (M=-105.1, SD=95.3) and 
cluttered (M=-101.0, SD=95.9) after the pairwise comparison using Bonferroni adjustments (p = .014). There was no significant difference between uncluttered and 
semi-cluttered (M=-102.0, SD=96.7) (p = .095), and cluttered and semi-cluttered (p = .095).

\subsubsection{Distances are Underestimated with narrower FOVs}

A significant effect of FOV was perceived through our findings (p<.001). Distance underestimation was the lowest and most accurate through the 165° $\times$ 110° view which is the widest FOV.
Moreover, the smallest FOV resulted in the highest underestimation of distance and the highest recorded error. In addition, distance judgment deteriorated with the decrease in the FOV. Figure~\ref{fig:MEDifferentFOVs} shows the error by FOV. 
We administered a pairwise comparison with Bonferroni adjustments which reflected a significant difference between the 
165° $\times$ 110° (M=-94.0, SD=95.5) and 
110° $\times$ 110° (M=-105.3, SD=96.6) views (p < .001) and also between 165° $\times$ 110° and 
45° $\times$ 35° (M=-107.7, SD=95.2) (p < .001). 
However, the difference was not significant between 110° $\times$ 110° and 45° $\times$ 35° (p = .085). Masnadi et al. also performed the same analysis and our results confirm theirs~\cite{masnadi2022effects}.

\subsubsection{Indoor vs Outdoor}
We found a significant difference between indoor (M=-71.4, SD=80.3) and outdoor (M=-134.0, SD=100.0) environments (p<0.001). This is inline with previous studies in the literature. Figure~\ref{fig:MEbyDistancebyEnv} shows the mean error for each target distance by environment

\subsubsection{FOV and Clutter have an independent effect on distance perception}
No significant interaction between clutter and FOV was perceived, which reflects the fact that FOV and clutter independently impact the distance judgment. 
The improvement of distance perception when clutter and FOV increase can be observed in Figure~\ref{fig:MEbyClutterbyFOV}.

\section{Discussion}

Based on the performed blind-walk user study, we have found that the presence of clutter in VEs results in the improvement of distance judgment and perception. Along with that, the study results show that wider FOVs cause more accurate distance perception and less underestimation.

%
%
\subsection{H1: Clutter in Virtual Environments}
Based on the study results, it is perceivable that the increase of clutter in VEs leads to a better perception of distance along with decreasing the underestimation of distance. In the conducted study, we experimented using three different levels of clutter for each of the four environments that we used, this serves as a means to be able to generalize our idea that the increase of clutter in an environment results in an improvement of the perception of distance.
This finding reveals that distance perception is improved after adding objects to the scene and also by adding more clutter to it independent from any influence of being in an indoor or outdoor setting.

\subsection{H2: Indoor vs Outdoor Environments}
Through the performed study, we perceived that participants showed better performance in indoor virtual settings compared to outdoor ones. We assume that this might be related to the participants being more used to indoor environments and mostly due to the fact that they are accustomed to home alike settings along with them performing more daily tasks in indoor environments. This difference still needs additional investigation in order to better understand the underlying factors that differentiate indoor and outdoor environments.

\subsection{H3: FOV of the HMD}
Three FOVs were used throughout the investigation (165° $\times$ 110°, 110° $\times$ 110°, and 45° $\times$ 35°). Based on the findings through the study, a significant difference between the widest FOV and the two other FOVs was found, which can be justified by the fact that the stimulation of the far-periphery area of the eye which is above 120° is only possible through the 165° $\times$ 110° FOV.

\subsection{H4: Clutter and FOV}
We found that clutter and FOV had an independent effect on the perception of distance since no interaction effect was perceived. This shows that regardless of FOV, clutter can improve the perception of distance and the same applies to FOV regardless of clutter level.

\section{Limitations and Future Work}
The results of our study can be mainly used in a VR context and the same results cannot be claimed to be applicable in AR or real-world as further investigation is required with the aim of generalizing our results. Therefore, it is crucial to perform studies to evaluate the effects of clutter on egocentric distance perception through AR devices and also in real world.

In addition, generalizing the results found to all potential VR users population is still questionable as the targeted population of the study was mainly from the student body of the university, which had limitation of the age of the participants and as a consequence contributed to a limitation of the findings of our study.

There was not a direct comparison with real-world environments. Our conducted study was focused on the VEs and their characteristics along with comparing them to each other besides evaluating the impact of clutter in each one of those environments. Our effort was to diversify the possible VEs in order to be able to generalize the findings to various VEs.

In this paper, we defined clutter as the number of objects present in the scene of a chosen environment. Clutter can have different definitions in different contexts, which can be exemplified through the following: the overall contrast of the scene, the total length of the edges, and also set the clutter definition based on the information theory and so forth, which represent different approaches to set the definition of clutter for future investigations.

We stated beforehand, users tend to perform better in indoor environments, which will require additional investigation with the aim of justifying this phenomenon. This probably correlates to the familiarity with the indoor environments and users being accustomed to the size of the indoor objects. Moreover, there are walls and ceilings in an indoor environment that can provide perspective cues to the user.

In our study, users could move their eyes around while looking at a limited FOV which resulted in stimulating the peripheral vision with depth cues, even in narrower FOVs. In future work, adopting headsets with eye-tracking systems would make it possible to better understand the periphery and far-periphery stimulation effect on the perception of distance. In such a system, the limited FOV can follow the user's gaze and makes it possible to isolate periphery stimulation.

\section{Conclusion}
We performed a study using a blind-walking task that represents an action-based evaluation. We found improvement in distance judgment accuracy and a reduction of distance underestimation by adding more clutter and objects to the environment. Based on our results, using cluttered environments improves the perception of distance in contrast with environments without clutter. In addition, we have shown that this improvement of distance judgment through adding more clutter was independent of FOV. In this study, we consider the environmental characteristics to be significant, considering that results showed participants perform better indoors in the blind-walking task, whereas more underestimation was recorded in the outdoor cases. Based on these findings, it is important to mention that such environmental factors should be highlighted and emphasized on when developing VR-based environments, tasks, and systems.

\begin{acks}
We would like to thank all the members that contributed to this project and the anonymous reviewers for their valuable feedback.
\end{acks}
\bibliographystyle{ACM-Reference-Format}
\bibliography{MAIN}

\end{document}